\documentclass{aa}  
\usepackage{hyperref}
\usepackage{natbib}
\usepackage{ulem}
\usepackage{xcolor}
\usepackage{graphicx}
\usepackage{txfonts}
\begin{document} 

\newcommand{\Tcool}{$3.3_{-0.4}^{+0.3}\,$MK}
\newcommand{\Thot}{$13.1_{-1.4}^{+2.0}\,$MK}
\newcommand{\eom}{$(3.7\pm0.7)\times 10^{30}\,$erg}
\newcommand{\eepic}{$(4.7\pm2.2)\times 10^{30}\,$erg}
\newcommand{\xraydur}{$4.5\,$ks}
\newcommand{\Tqmean}{$6.5\,$MK}
\newcommand{\Tfmean}{$11.3\,$MK}
\newcommand{\FX}{$(6.3 \pm 0.4)\times 10^{-14} \,\rm{erg}\, \rm{cm}^{-2} \rm{s}^{-1}$}
\newcommand{\LXquiet}{$(4.56 \pm 0.62)\times 10^{26}\,\rm{erg}\,\rm{s}^{-1}$}
\newcommand{\LXflaring}{$(0.46 \pm 0.06)\times 10^{27}\,\rm{erg}\,\rm{s}^{-1}$}
\newcommand{\Lbol}{$(5.0 \pm 0.3) \times 10^{30}\,\rm{erg}\,\rm{s}^{-1}$}
\newcommand{\LXLbol}{$(0.9\pm 0.1)\times 10^{-4}$ }
\newcommand{\ffdalpha}{$-1.86_{-0.21}^{+0.18}$}
\newcommand{\ffdr}{$-0.94\,\mathrm{d}^{-1}$}

   \title{The corona of a fully convective star with a near-polar flare}

   \author{E. Ilin\inst{1,2},
            K. Poppenh\"ager\inst{1,3},
            B. Stelzer\inst{4} and
            D. Dsouza\inst{1,3}
          }

   \institute{Leibniz Institute for Astrophysics Potsdam (AIP), An der Sternwarte 16, 14482 Potsdam, Germany\and
              ASTRON, Netherlands Institute for Radio Astronomy, Oude Hoogeveensedijk 4, Dwingeloo, 7991 PD, The Netherlands\\
                            \email{ilin@astron.nl}\and
    Institute for Physics and Astronomy, University of Potsdam, Karl-Liebknecht-Strasse 24/25, 14476 Potsdam, Germany \and              
                    Institut für Astronomie \& Astrophysik, Eberhard Karls Universität Tübingen, Sand 1, 72076 Tübingen, Germany        }

   \date{Received Feb 8, 2024; accepted May 6, 2024}

 
  \abstract
{In 2020, the \textit{Transiting Exoplanet Survey Satellite} (\textit{TESS}) observed a rapidly rotating M7 dwarf, TIC~277539431, produce a flare at $81^{\circ}$ latitude, the highest latitude flare located to date. This is in stark contrast to solar flares that occur much closer to the equator, typically below $30^{\circ}$. The mechanisms that allow flares at high latitudes to occur are poorly understood.} 
{We studied five Sectors of \textit{TESS} monitoring, and obtained 36 ks of \textit{XMM-Newton} observations to investigate the coronal and flaring activity of TIC~277539431.}
{From the observations, we infer the optical flare frequency distribution, flare loop sizes and magnetic field strengths, the soft X-ray flux, luminosity and coronal temperatures, as well as the energy, loop size and field strength of a large flare in the \textit{XMM-Newton} observations.}
{We find that TIC~277539431's corona does not differ significantly from other low mass stars on the canonical saturated activity branch with respect to coronal temperatures and flaring activity, but shows lower luminosity in soft X-ray emission by about an order of magnitude, consistent with other late M dwarfs.} 
{The lack of X-ray flux, the high latitude flare, the star's viewing geometry, and the otherwise typical stellar corona taken together can be explained by the migration of flux emergence to the poles in rapid rotators like TIC~277539431 that drain the star's equatorial regions of magnetic flux, but preserve its ability to produce powerful flares.}
   {}


   \maketitle
%

\section{Introduction}
\label{sec:intro}
Most M dwarfs hosts terrestrial planets, with a sizable fraction orbiting at instellations where liquid water can exist~\citep{dressing2015occurrence, hardegree-ullman2019kepler, ment2023occurrence}. However, to be habitable in an Earth-like manner, merely the possibility of surface water in its liquid form does not suffice. Space weather of the host, that is, the high energy radiation and particles that impact the planet's atmosphere, affect its hospitality for life, particularly for M dwarfs~\citep{airapetian2020impact}. If the energetic photon or particle flux is too high, the atmosphere may blow off~\citep[e.g., ][]{lammer2003atmospheric, garraffo2017threatening, ketzer2022influence} and water may be lost from the surface~\citep{doamaral2022contribution}. If too low, life may not emerge in the first place~\citep{rimmer2018origin}.

In M dwarfs, stellar activity evolution on the main sequence unfolds much more slowly than in FGK stars. For the latter, magnetic activity, and with it the flares, winds, coronal mass ejections and energetic particle eruptions that make up their space weather, decline rapidly over a few hundred megayears. A fully convective M dwarf stays highly active for gigayears~\citep{magaudda2020relation, johnstone2021active, medina2022galactic}. Moreover, the zone where liquid water can reside is much closer to these faint stars. The result is a planet that may be exposed to violent space weather conditions for a good fraction of the age of the universe. 

Stellar space weather originates from the stellar corona. The transition from star to brown dwarf is characterized by increasing rotation speed and decreasing coronal emission. At the same time, the observed surge of radio emission at the bottom of the main sequence indicates a transition from stellar corona to planet-like magnetosphere~\citep{zarka1998auroral,pineda2017panchromatic}. Yet, brown dwarfs are regularly detected with energetic flares regardless of their expected low X-ray luminosity~\citep[e.g., ][]{ stelzer2006simultaneous, gizis2013kepler, paudel2020k2, schmidt2019largest, audard2007chandra, deluca2020extras}. How these flares are produced in the apparent absence of a solar-type corona is unclear~\citep{mullan2018frequencies}.

In \citet{ilin2021giant}, we searched fully convective late M dwarfs that mark the transition from star to brown dwarf with \textit{TESS}~\citep{ricker2015transiting}. We found four flares on four rapidly rotating ($P_{\rm rot}<10\,$h) M5-M7 dwarfs that lasted for multiple rotation periods of each star. The flares were observed to rotate in and out of view. The shape of this modulation together with the known inclination (combining $P_{\rm rot}$ and $v \sin i$ from high resolution spectroscopy) allowed us to determine the latitudes of these flares. These flares were found significantly closer to the pole than to the equator, in contrast to the Sun, where flares are usually found below $30^{\circ}$ latitude. Our results indicate a preference for flares at high latitudes in these stars, as a chance finding of these latitudes among latitudinally equidistributed flares was unlikely ($\sim 0.1\%$). If flares and associated particle eruptions indeed had a latitude preference, the space weather of a planet in an aligned orbit would be less severe than previously thought. 


Among the studied objects, TIC~277539431, TIC~277 for short, showed the highest latitude flare known to date at $\sim81^{\circ}$~\citep[][Table \ref{tab:starparams}]{ilin2021giant}. It is an M7 dwarf with an extremely short rotation period of $4.56\,$h. While the fast rotation and late spectral type could be indicative of a diminishing corona, the detection of flares suggests otherwise.

In this work, we follow TIC~277 with \textit{XMM-Newton} for $36\,$ks and use optical monitoring from \textit{TESS}~(Section~\ref{sec:obs}) to measure its coronal and flaring properties, respectively~(Section~\ref{sec:methods}). We contrast our results~(Section~\ref{sec:results}) with the literature to investigate if TIC~277 behaves like a low mass star or brown dwarf, consider scenarios that can explain the combined observations~(Section~\ref{sec:discussion}), and summarize our results in Section~\ref{sec:summary}.


\begin{table}
\footnotesize
\centering
    \caption{Stellar properties of TIC~277539431.}
    \begin{tabular}{ll}\hline 
         Parameter & Value  \\\hline
         
         Alternative ID & WISEA J105515.71-735611.3 \\
         Spectral type & M7 [1]\\
         Distance $d$ & $13.70\pm0.11\,$pc [2] \\
         Effective temperature $T_{\rm eff}$ & $2680^{60}_{-50}\,$ K [3]\\
         Rotation period $P_{\rm rot}$ & $273.618 \pm 0.007\,$min [1]\\
         Projected rot. velocity $v\sin i$ & $38.6\pm1.0\,$km/s [1] \\
         Inclination $i$ & $(87.0^{+2.0}_{-2.4})^{\circ}$ [1]\\
         Radius $R$ & $0.145\pm0.004\,R_\odot$ [1]\\
        
    \end{tabular}
    \newline\footnotesize
    \flushleft
    [1] \citet{ilin2021giant}, 
    [2] \citet{bailer-jones2018estimating}, 
    [3] \citet{pecaut2013intrinsic}
    \label{tab:starparams}
\end{table}

\section{Observations}
\label{sec:obs}

\subsection{XMM-Newton}
\label{sec:obs:xmm}
\textit{XMM-Newton} is an X-ray telescope that was launched into orbit in 1999. It was equipped with six science instruments that operate simultaneously, among them the European Photon Imaging Camera with three cameras, PN, MOS1 and MOS2; and the Optical Monitor (OM) that provides near-UV and optical photometry, among other capabilities.
We used \textit{XMM-Newton} to observe TIC~277 for $36\,$ks, i.e. two full rotation periods of the star, on August 5, 2022 (Proposal 090120; PI: E. Ilin). We used the EPIC instruments aboard \textit{XMM-Newton}, i.e., MOS1/2 and PN, that cover the soft X-ray band from $0.2$ and up to $12\,$keV, as well as the OM using its white light filter.

\subsubsection{Time series with EPIC and OM}

\begin{figure*}[ht!]
    \begin{centering}
        \includegraphics[width=0.75\linewidth]{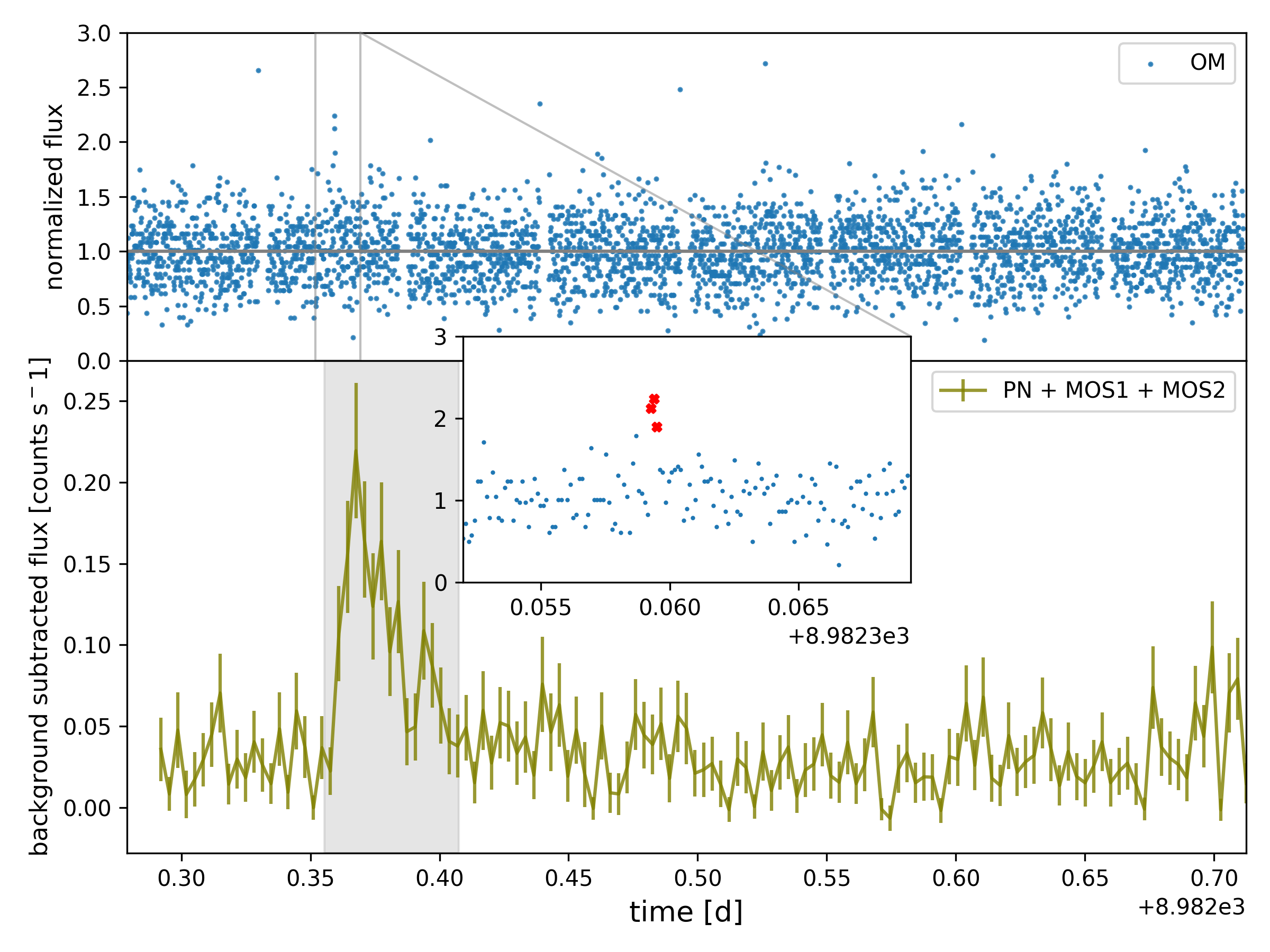}
        \caption{
         Top panel: \textit{XMM-Newton} Optical Monitor light curve. Bottom panel: Background-subtracted \textit{XMM-Newton} X-ray light curve using the flux in the entire $0.2-10\,$keV range  (PN, MOS1 and MOS2 combined). The grey shaded portion defines the flare-only subset of the observations (see Section~\ref{sec:methods:epic} and Table~\ref{tab:specfit}). The inset shows the marginal OM flare, with the flare data points marked with red crosses.
        }
        \label{fig:lightcurves}
    \end{centering}
\end{figure*}

Fig.~\ref{fig:lightcurves} shows the time series using the combined PN and MOS time series together with the OM light curve. We extracted the events time series from EPIC using XMM SAS version 20\footnote{"Users Guide to the \textit{XMM-Newton} Science Analysis System", Issue 18.0, 2023 (ESA: \textit{XMM-Newton} SOC)} using the \texttt{evselect} task in the full $0.2-10\,$keV band. We selected a circular source region with 20 arcsec radius on all detectors, and a circular source-free background region with 120 and 90 arcsec radius in MOS and PN, respectively. We used the \texttt{epiclccorr} task in XMM SAS to correct for energy and time dependent loss of events\footnote{\url{https://heasarc.gsfc.nasa.gov/docs/xmm/sas/help/epiclccorr/node3.html}, accessed on July 3, 2023}, and extracted a soft X-ray time series with a time binning of $200\,$s.

Simultaneously, we used the OM aboard \textit{XMM-Newton} to monitor TIC~277 in its white light filter at $10$\,s cadence. The OM white light filter is centered on $4060\,$\AA, with an equivalent width of $3470\,$\AA. In each light curve, OM observed uninterruptedly for $4390\,$s, followed by a short gap of $318\,$s before the next one. We used the \texttt{omfchain} in XMM SAS to extract the 8 individual light curves.

\subsubsection{EPIC Spectra}
For the EPIC observations, we used the standard spectral extraction procedure for point sources, as specified in the \textit{XMM-Newton} SAS handbook. We extracted spectra using \texttt{evselect} to select events in the circular event and background regions, \texttt{epatplot} to check for pile-up of multiple photons in one exposure, \texttt{backscale} to calculate the area of the source region, and \texttt{rmfgen} and \texttt{arfgen} to convolve the observed energies with the instrument response, and produce the spectra. Finally, we used \texttt{grppha} to rebin the data in the spectra to obtain at least 15 counts per bin. To produce spectra for the quiescent and flaring portions of the observations separately, we used \texttt{tabgtigen} to select the time intervals to pass to \texttt{evselect}.

\subsection{TESS}
\label{sec:obs:tess}
Since 2018, \textit{TESS} has been supplying publicly available red-optical high-precision time series photometry in an ongoing all-sky survey. Each uninterrupted observing Sector provides an approximately 27 d long light curve in a broad $6000$-$10000\,$\AA\,band pass. 

\begin{figure*}
    \begin{centering}
        \includegraphics[width=0.9\hsize]{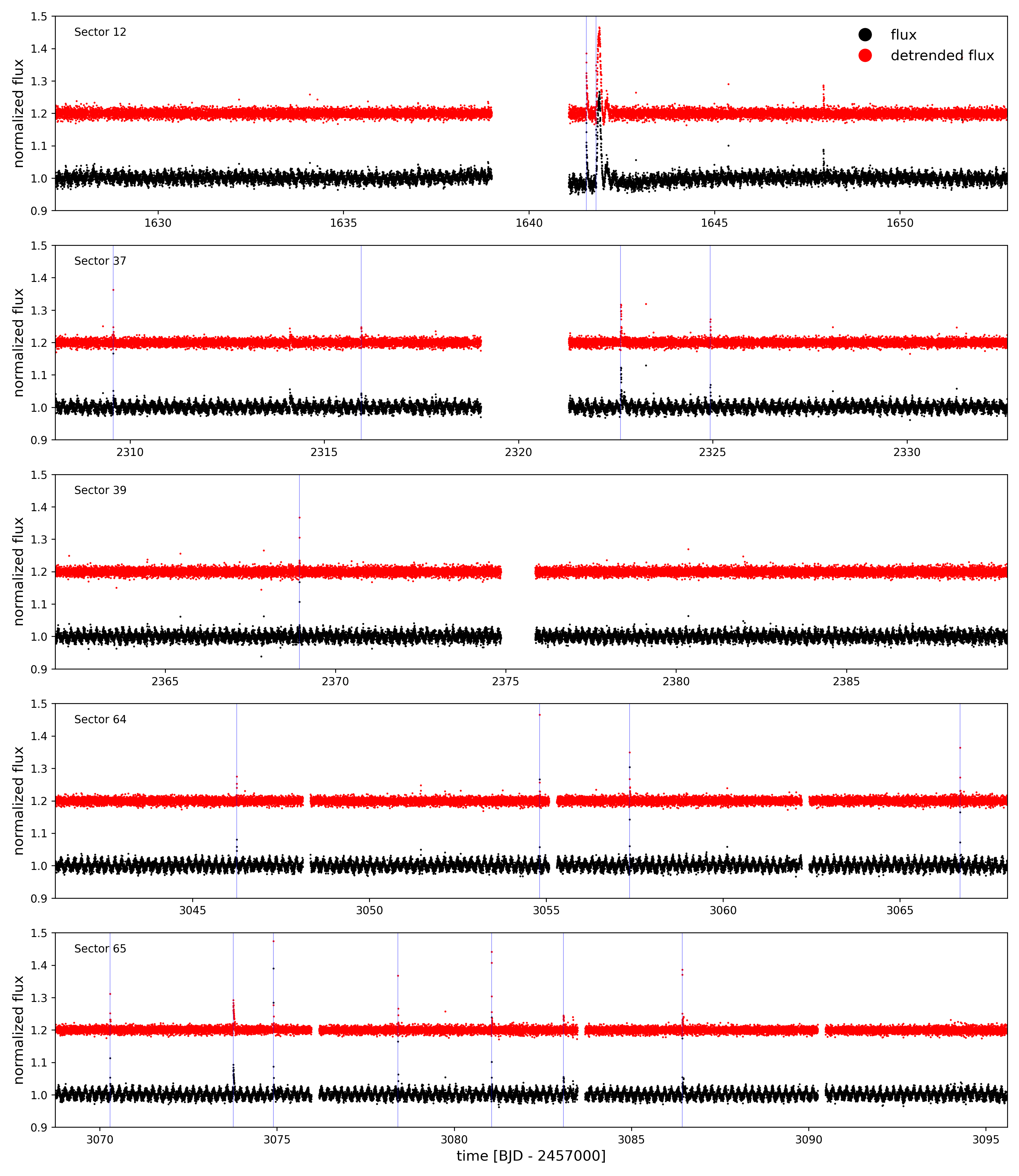}
        \caption{
          Normalized \textit{TESS} light curves. Black and red dots show the light curve with and without rotational variability and trends, respectively. Blue vertical lines mark the start times of flare candidates in Table~\ref{tab:flares}. The de-trended light curve is offset by $0.2$ for better visibility. The large, rotationally modulated, flare, localized at about $81^{\circ}$ latitude presented by~\citet{ilin2021giant}, appears in the second half of the top panel~(Sector 12).}
        \label{fig:tess_lcs}
    \end{centering}
\end{figure*}

We used all 5 Sectors of optical monitoring of TIC~277 at a $2\,$min cadence. Sector 12 was observed in May 2019, Sectors 37 and 39 in April and June 2021, and Sectors 64 and 65 in April and May 2023~(Fig.~\ref{fig:tess_lcs}).

\section{Methods}
\label{sec:methods}


In this work, we analyse the coronal properties and flaring behavior of the late M dwarf TIC~277. From the X-ray observations with \textit{XMM-Newton}, we obtain a soft X-ray spectrum, which we model with two thermal emitter components~(Section~\ref{sec:methods:epic}). In \textit{TESS}' optical time series photometry, we remove the variability introduced by the star's rotation, and find and characterize the flares in the de-trended light curves~(Section~\ref{sec:methods:tess}). \textit{XMM-Newton} captured a flare simultaneously in both the PN and MOS instruments in X-ray, and marginally the OM instrument in the optical~(Fig.~1 and Sections~\ref{sec:methods:om} and~\ref{sec:methods:epicflare}). We derive its soft X-ray energy using the luminosity derived from the X-ray spectrum, and its bolometric energy from the OM data. For the following comparison between optical and X-ray flaring activity, we also calculate the bolometric energies of the flares detected in \textit{TESS}~(Section~\ref{sec:methods:flareenergies}), and estimate the flaring loop size and magnetic field strength of both \textit{TESS} and \textit{XMM-Newton} flares~(Section~\ref{sec:methods:loops}).

\subsection{Spectral fitting: EPIC}
\label{sec:methods:epic}
The stellar corona can be described as an optically thin thermal plasma that consists of multiple temperature components that represent different regions. The number of components one can identify in an X-ray spectrum depends on the brightness of the source. We use XSPEC version 12.12.1~\citep{arnaud1996xspec} to fit a two-temperature (2T) additive VAPEC~\citep{smith2001collisional,foster2012updated} model to the joint PN and MOS spectra, using solar abundances from~\citet{grevesse1998standard}, but adjusting to an Fe/O$=0.6$ ratio, which is more typical of M dwarfs~\citep{wood2018chandra}. We analyzed both the full data set, and the quiescent and flaring parts separately~(see Fig.~\ref{fig:lightcurves} for the definition of the flare time interval). Neither subset could be adequately fit with a single temperature component (1T), and a 3T model did not improve the fit compared to the 2T-model. In the full data set, we used the \texttt{bayes} method in XSPEC to assign constant priors within the model's hard limits, and sample the uncertainties using the Markov-Chain-Monte-Carlo (MCMC) method using \texttt{chain} for a total of 30000 steps, discarding the first 5000 as the burn-in phase. From the same fit, but using PN data only, we derived the flux $F_X$ and X-ray luminosity $L_X$ in the soft $0.2-2\,$keV band. For the quiescent and flaring portions of the observations, we chose the same procedure, but used Gaussian prior distributions for the two coronal temperatures from the full data set, where we assigned the largest difference between the 50th, and 16th and 84th percentiles as standard deviation.

\subsection{Light curve de-trending and flare finding: TESS}
\label{sec:methods:tess}
We used AltaiPony~\citep{ilin2021altaipony} to de-trend the Pre-search Data Conditioning Simple Aperture Photometry (PDCSAP) light curves (Fig.~\ref{fig:tess_lcs}) using the de-trending function \texttt{custom\_detrending}\footnote{see also the documentation online at \url{https://altaipony.readthedocs.io/en/latest/tutorials/detrend.html}, accessed on July 3, 2023}, detailed in~\citet{ilin2022searching}. The procedure begins with a 3rd order spline fitting of coarse trends, followed by iterative sine fitting to remove rotational modulation, and two Savitzky-Golay filters~\citep{savitzky1964smoothing} with a $6\,$h and $3\,$h window, respectively to remove aperiodic short term trends, while masking potential flare candidates in each step. The masking takes three steps each time. First, all outliers above $>2.5\sigma$ are masked. Second, to mask the flare decay, we append the six-fold number of data points masked in the first step to the mask. Third, three adjacent data points before and after the so expanded mask are added to the mask. 

In the de-trended light curves, we searched for flare candidates as at least 3 consecutive $>3\sigma$ outliers above the noise level. Here the noise level is defined as the rolling standard deviation of the de-trended light curve flux within a $2$\,h window, filling in the noise level inside the previously masked areas with the mean of the noise levels adjacent to the mask. To the series of outliers, we kept appending data points to the flare candidate until a data point falls below $2\sigma$. 

\subsection{Light curve de-trending and flare finding: OM}
\label{sec:methods:om}

For the OM data, we performed a simplified analysis in order to identify flare candidates. The photometric noise in the OM data is much higher than in the \textit{TESS} light curves, so that the rotational modulation is drowned out. We therefore adopted the median value of each light curve, and searched for series of at least three consecutive positive outliers $3\sigma$ above the median. We find one candidate with three data points, shown in the inset in Fig.~\ref{fig:lightcurves}. We call this flare candidate marginal because it only barely passes the threshold. The candidate has in fact a relative amplitude of $1.22$, larger than any flare detected in the \textit{TESS} data, but is also much more impulsive, lasting less than a minute, below the 2-min resolution of \textit{TESS}. While the evidence for this event from the OM alone is weak, its timing $12\,$min before the peak of an X-ray flare lends credibility to this detection~(see Section~\ref{sec:results:flares}).

\subsection{EPIC flare detection}
\label{sec:methods:epicflare}

In the EPIC light curve, we applied the same approach to finding flares as for OM~(Section~\ref{sec:methods:om}), because we did not find any rotational modulation in EPIC either. We found one flaring event that stood out. In addition to the data points $3\sigma$ or more above the median level, we count several data points towards the flare before and after to make sure that flaring and quiescent portions of the light curve are clearly separated (see shaded region in Fig.~\ref{fig:lightcurves}). 

\subsection{Flare energies: TESS, OM, and EPIC}
\label{sec:methods:flareenergies}
The OM data contain one marginal flare that we interpret as associated with the large X-ray flare shortly after~(inset in Fig.~\ref{fig:lightcurves}). In the TESS light curves, we found a total of 18 flares. To assess the flaring activity of TIC~277, we computed the bolometric energies for both the OM and the \textit{TESS} flares. 

For the OM flare, we used the median flux of the light curve as the quiescent flux $F_0$, against which we measured the equivalent duration ($ED$) of the flare, i.e., the flare flux $F_{\rm flare}$, divided by $F_0$, integrated over the flare duration~\citep{gershberg1972results}:

\begin{equation}
\label{eq:ED}
ED=\displaystyle \int \mathrm dt\, \frac{F_{\rm flare}(t)}{F_0}.
\end{equation}

We then converted the $ED$ to flare energy using the bolometric flare energy following the procedure in~\citet{shibayama2013superflares}. We used the throughput curve for the white light filter scaled to unity at the peak of transmission as an optimistic response curve $R_{\lambda, \rm OM}$, given that the degradation of this filter's response is poorly constrained\footnote{\textit{XMM-Newton} User's Handbook 3.5.3.1, \url{https://xmm-tools.cosmos.esa.int/external/xmm_user_support/documentation/uhb/omfilters.html}, accessed Aug 8, 2023}. From that, we extracted a ratio $f_{\rm OM}$ of stellar to flare flux of $1.4\cdot 10^{-4}$ for a $10000\,$K blackbody flare:

\begin{equation}
    f_{\rm OM} = \frac{\displaystyle\int R_{\lambda, \rm OM} B_{\lambda(T_{\rm eff})}  d\lambda}{\displaystyle\int R_{\lambda, \rm OM} B_{\lambda(T_{\rm flare})} d\lambda} 
    \label{eq:ratio}
\end{equation}

A flare temperature of $T_{\rm flare}=10000\,$K is a typical approximation for energetic M dwarf flares~\citep{kowalski2013timeresolved, howard2020evryflarea}, but deviations towards both hotter and cooler temperatures have been observed in the past, which further increases the uncertainty on our energy estimate~\citep[][and references therein]{kowalski2024stellar}. With the ratio $f_{\rm OM}$, we can calculate the bolometric flare energy as:

\begin{equation}
    E_{\rm flare} = ED \cdot \pi R^2 \cdot \sigma_{\text{B}} T_{\rm eff}^4 \cdot f_{\rm OM},
    \label{eq:eflare}
\end{equation}

where $\sigma_{\text{B}}$ is the Stefan-Boltzmann constant. We note that due to the optimistic response curve assumption, the resulting $E_{\rm flare}$ is likely higher in reality.


Analogously to Eq.~\ref{eq:ratio} for the OM flare, we used the \textit{TESS} response curve to obtain a flux ratio $f_{\rm TESS}$ of $6.3\cdot10^{-3}$ in the optical filter of \textit{TESS}, and convert $ED$ to bolometric flare energy using Eq.~\ref{eq:eflare}. Besides the $ED$, \texttt{AltaiPony} also yields start and end time, defined as the first and last data point above the flare detection criterion defined in Section~\ref{sec:obs:tess}; and relative amplitude $a$, for each flare.

For the X-ray flare energy, we multiply flare duration $\Delta t$, that is, the difference between start and end time, by the flare luminosity in the EPIC data. The flare luminosity is equal to the difference between the quiescent and flaring X-ray luminosities in the $0.2-2.0\,$keV band, i.e.:

\begin{equation}
    E_{X, \mathrm{flare}} = \left(L_{X,\mathrm{flaring}} -  L_{X,\mathrm{quiescent}}\right) \cdot \Delta t
    \label{eq:xrayflare}
\end{equation}

\subsection{Flare loop sizes and magnetic field strengths}
\label{sec:methods:loops}

Lacking spatial resolution of the stellar disk, we cannot directly measure the flare loop sizes and magnetic field strengths from either X-ray or optical observations. However, we can apply scaling relations based on solar observations~\citep{shibata2002hertzsprungrusselllike, namekata2017validation} that use X-ray and optical flare diagnostics as proxies.

\citet{shibata2002hertzsprungrusselllike} derived scaling relations for the magnetic field strength and loop size from solar soft X-ray observations (their Eqns.~7a and 7b):

\begin{equation}
    B = 50 \left( \frac{EM}{10^{48} \, \text{cm}^{-3}} \right)^{-0.2} \left( \frac{n_0}{10^9 \, \text{cm}^{-3}} \right)^{0.3} \left( \frac{T}{10^7 \, \text{K}} \right)^{1.7}\text{[G]}
    \label{eq:shibataB}
\end{equation}

\begin{equation}
    L = 10^9 \left( \frac{EM}{10^{48} \, \text{cm}^{-3}} \right)^{0.6} \left( \frac{n_0}{10^9 \, \text{cm}^{-3}} \right)^{-0.4} \left( \frac{T}{10^7 \, \text{K}} \right)^{-1.6} \text{[cm]}
    \label{eq:shibataL}
\end{equation}



We also follow~\citet{maehara2021timeresolved}, who estimate the magnetic field strengths and flaring loop sizes for the OM and \textit{TESS} flares based on the relation between flare \textit{e}-folding time $\tau$ and bolometric energy $E_{\rm flare}$ derived from the magnetic reconnection model, and calibrated on solar flares in~\citet{namekata2017validation}, who themselves use the scaling relations in~\citet{shibata2002hertzsprungrusselllike}. \citet{namekata2017validation} argue that the \textit{e}-folding time $\tau$ is similar to the magnetic field reconnection time, and that the flare energy $E_{\rm flare}$ is similar to the magnetic energy stored in the active region where the flare occurs. Using their Eqns. 9 and 10, we obtain: 

\begin{equation}
    \tau = 3.5 \left(\dfrac{E_{\text{flare}}}{1.5\cdot10^{30}\,\text{erg}}\right)^{1/3} \left(\dfrac{B}{57\,\text{G}}\right)^{-5/3} \,[\text{min}]
    \label{eq:maeharaB}
\end{equation}

Using the approximation that the flare energy is proportional to the magnetic energy stored in the flaring volume, i.e., $E_{\text{flare}}\propto B^2 L^3$~\citep{shibata2013can}, we can rephrase the $\tau-E_{\text{flare}}$ relation in Eq.~\ref{eq:maeharaB} in terms of the size of the flaring region, which should be of the order of the loop size, i.e.:

\begin{equation}
    \tau = 3.5 \left(\dfrac{E_{\text{flare}}}{1.5\cdot10^{30}\,\text{erg}}\right)^{-1/2} \left(\dfrac{L}{2.4\cdot10^{9}\,\text{cm}}\right)^{-5/2}\,[\text{min}]
    \label{eq:maeharaL}
\end{equation}

\section{Results}
\label{sec:results}
In this work, we use the \textit{XMM-Newton} and \textit{TESS} observations to constrain the coronal and flaring properties of TIC~277. We derive its coronal temperature and luminosity from the EPIC spectrum~(Section~\ref{sec:res:XrayTL}). We also measure the flare energies in X-rays from the EPIC instruments, and in optical from the OM, and compare them to the flare frequency distribution obtained from the 18 detected flares in \textit{TESS}~(Section~\ref{sec:results:flares}). Finally, we estimate the flaring loop sizes and magnetic field strengths of the \textit{XMM-Newton} and \textit{TESS} flares using scaling relations based on solar observations~(Section~\ref{sec:results:loops}).

\subsection{Coronal temperature and luminosity}
\label{sec:res:XrayTL}

\begin{figure}
    \begin{centering}
        \includegraphics[width=\linewidth]{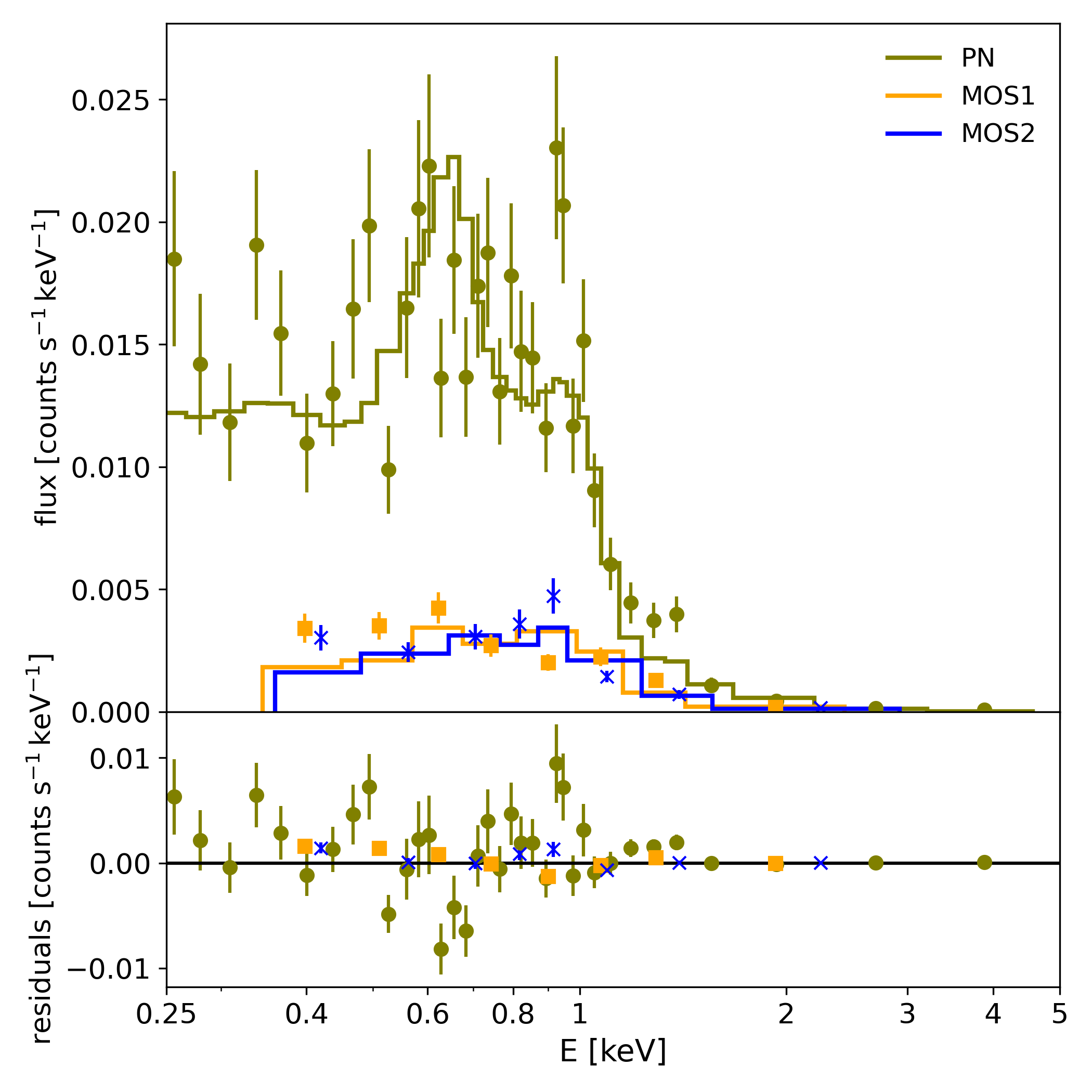}
        \caption{
         Time-averaged \textit{XMM-Newton} EPIC spectra. The top panel shows the spectra taken with the different EPIC instruments (data points with error bars) together with the best-fit two-temperature VAPEC model with Fe/O$=0.6$ for the full data set (solid lines). The bottom panel shows the residuals to the fit. 
        }
        \label{fig:spec_joint_all}
    \end{centering}
\end{figure}

From the \textit{XMM-Newton} EPIC spectra, we measured the coronal properties of TIC~277 in both its quiescent, and flaring states~(Table~\ref{tab:specfit}). The resulting fit for the full data set~(Fig.~\ref{fig:spec_joint_all}) consists of a cooler and a hotter component of about \Tcool\,and \Thot, respectively. Assuming these temperatures for the prior distribution, we also fitted 2T VAPEC models to the flaring and quiescent portions of the observations separately. The coronal temperatures did not change significantly using either subset. However, the emission measure weighted mean temperature $T_{\rm mean}$ changed from \Tqmean\,in the quiescent spectrum to \Tfmean\,in the flaring spectrum, when the hot component became dominant during the flare. From the quiescent PN spectra, we then also calculated the X-ray flux of $F_X=$\FX, and luminosity $L_X=$\LXquiet\,in the soft $0.2-2.0\,$keV band. 

We note that not using the full set of observations as the prior resulted in an unconstrained hot coronal component during the flare, which could be due to a lack of X-ray flux above $5\,$keV. 

 \begin{table}
\footnotesize
\centering
    \caption{XSPEC fits to EPIC spectra for different subsets of observations. Fluxes and luminosities are given in the $0.2-2\,$keV band.}
\begin{tabular}{llll}
\hline
 & full data set & quiescent & flaring \\
\hline
$L_X$ [$10^{26}$ erg s$^{-1}$] & $5.5 [0.3]$ & $4.6 [0.6]$ & $14.1 [1.0]$ \\
$F_X$ [$10^{-14}$ erg s$^{-1}$ cm$^{-2}$] & $2.5 [0.1]$ & $2.0 [0.3]$ & $6.3 [0.4]$ \\
$T_{\rm cool}$ [MK] & $3.3^{0.3}_{-0.4}$ & $3.2^{0.2}_{-0.3}$ & $3.4^{0.3}_{-0.4}$ \\
$T_{\rm hot}$ [MK] & $13.1^{2.0}_{-1.4}$ & $12.4^{1.8}_{-1.5}$ & $13.1^{1.6}_{-1.3}$ \\
$10^6$ norm$_{\rm cool}$ & $8.0^{1.0}_{-1.1}$ & $7.2^{0.8}_{-0.8}$ & $7.5^{3.1}_{-3.2}$ \\
$10^6$ norm$_{\rm hot}$ & $6.6^{0.9}_{-0.9}$ & $3.9^{0.8}_{-0.7}$ & $32.5^{6.5}_{-5.5}$ \\
$\log_{10} EM_{\rm cool}$ [cm$^{-3}$] & $49.3[0.1]$ & $49.2[0.1]$ & $49.2[0.2]$ \\
$\log_{10} EM_{\rm hot}$ [cm$^{-3}$] & $49.2[0.1]$ & $48.9[0.1]$ & $49.9[0.1]$ \\
$T_{\rm mean}$ [MK] & $7.8[0.9]$ & $6.5[0.8]$ & $11.3[1.4]$ \\
\hline
\end{tabular}

        \label{tab:specfit}
\end{table}

\subsection{Flares}
\label{sec:results:flares}

\begin{table}
\centering
    \caption{Flares detected with \textit{TESS}. $t_s$ is the starting time of the flare, $a$ is its relative amplitude, and $E_{\rm flare}$ is the bolometric energy assuming a $10000\,$K blackbody emission from the flare.}
   \begin{tabular}{rrlr}
\hline
$t_{s}$ [BJD - 2457000] & $a$ & $E_{\rm flare}$ [$10^{31}$ erg] & Sector \\
\hline
1641.543564 & 0.185181 & 26.2 [0.5] & 12 \\
1641.811615 & 0.265880 & 257.6 [1.5] & 12 \\
2309.566699 & 0.162635 & 3.7 [0.2] & 37 \\
2315.952947 & 0.047461 & 2.3 [0.2] & 37 \\
2322.628060 & 0.117076 & 15.3 [0.4] & 37 \\
2324.937816 & 0.071603 & 4.4 [0.3] & 37 \\
2368.939146 & 0.167062 & 4.9 [0.2] & 39 \\
3046.245679 & 0.074594 & 2.5 [0.2] & 64 \\
3054.812486 & 0.266024 & 5.7 [0.2] & 64 \\
3057.356968 & 0.309830 & 10.4 [0.2] & 64 \\
3066.700102 & 0.163675 & 3.8 [0.1] & 64 \\
3070.290403 & 0.111329 & 4.7 [0.2] & 65 \\
3073.764022 & 0.091546 & 15.7 [0.5] & 65 \\
3074.900134 & 0.378347 & 11.0 [0.1] & 65 \\
3078.414016 & 0.167239 & 5.1 [0.2] & 65 \\
3081.055673 & 0.240892 & 11.0 [0.2] & 65 \\
3083.080666 & 0.044786 & 4.4 [0.3] & 65 \\
3086.430649 & 0.185751 & 8.7 [0.2] & 65 \\
\hline
\end{tabular}

        \label{tab:flares}
\end{table}

In \textit{TESS}, we found a total of 18 flares in the five Sectors, or equivalently, 125 days of observing at 2-min cadence. We fit the flare frequency distribution (FFD) with a power law of the form

\begin{equation}
    f(E) \mathrm{d} E = \beta \cdot E^{-\alpha} \mathrm{d} E.
\end{equation}
following the procedure in~\citet{ilin2021flares}, implemented as \texttt{FFD.fit\_powerlaw} using the \textit{mcmc} argument in \texttt{AltaiPony}~\citep{ilin2021altaipony}, which is using the posterior distribution derived in \citet{wheatland2004bayesian}, and sampling the posterior distribution with the \texttt{emcee} package~\citep{foreman-mackey2013emcee}.

The FFD power law fit converged after 13,500 steps on a slope $\alpha =$\ffdalpha\unskip. The frequency $R_{31.5}$, that is the frequency of flares per day above $\log_{10} E_{\rm flare}\,[\mathrm{erg}] = 31.5$ is $\log_{10}R_{31.5}=$\ffdr\unskip. The slope is typical of other flaring stars, regardless of spectral type and rotation period~(see, e.g., Fig.~13 of \citealt{ilin2021flares}). $R_{31.5}$ is typical of flaring M dwarfs, including late M dwarfs in the saturated activity regime, where $R_{31.5}$ becomes independent of rotation period~\citep{medina2020flare,murray2022study}. 

\begin{figure}
    \begin{centering}
        \includegraphics[width=\linewidth]{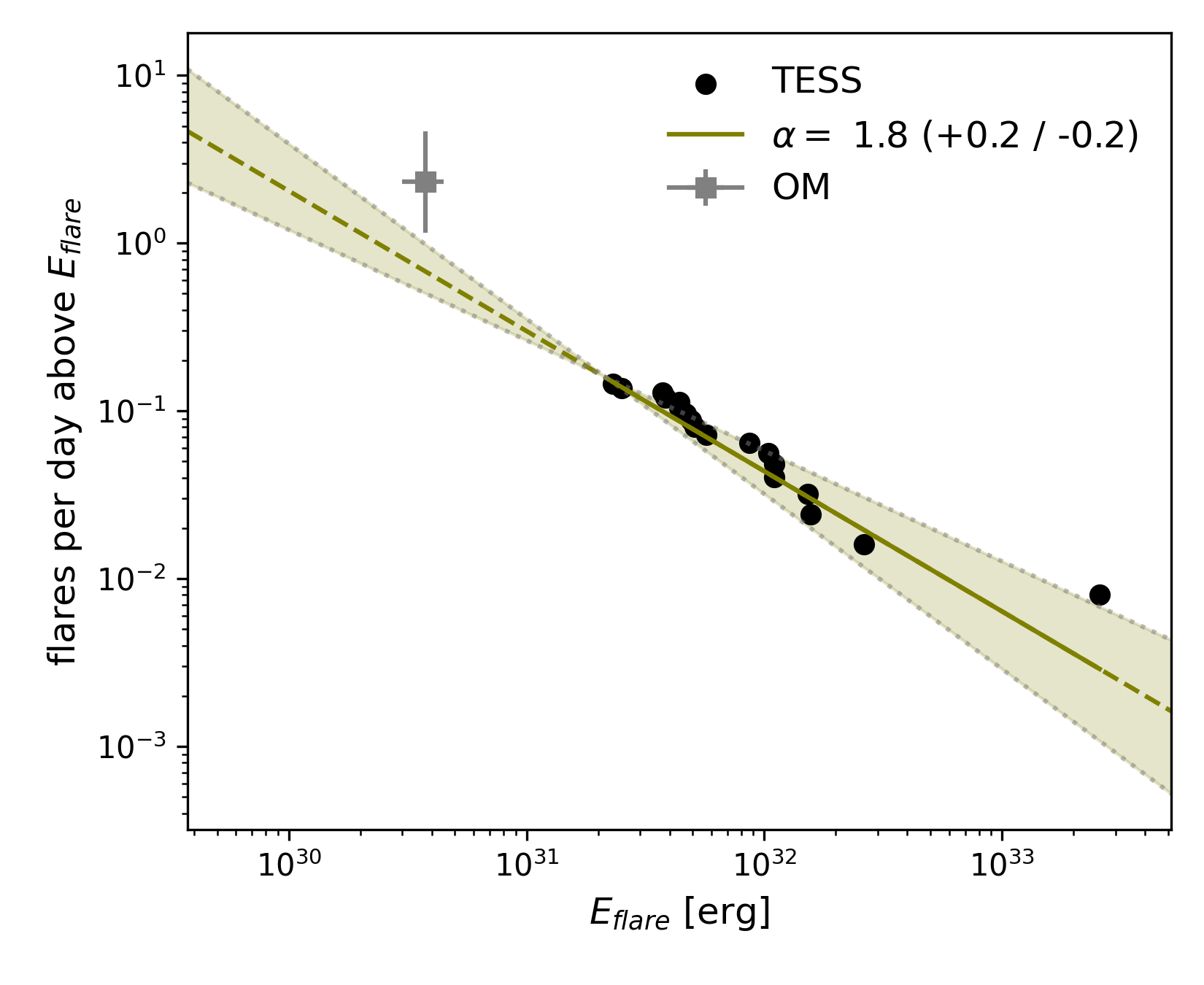}
        \caption{
         Cumulative flare frequency distribution of \textit{TESS} flares (black dots), with a power law fit (green line), and uncertainties (dotted lines). The rate of OM flares (gray square) is calculated from the observing baseline of \textit{XMM-Newton}. The flare energies are corrected for the pass bands of \textit{TESS} and OM, i.e., given as bolometric flare energies~(see Section~\ref{sec:methods:flareenergies}). 
        }
        \label{fig:ffd}
    \end{centering}
\end{figure}

For the X-ray flare in Fig.~\ref{fig:lightcurves}, we obtain a total energy of \eepic\;in the $0.2-2.0\,$keV band by inserting the quiescent and flaring luminosities (Table~\ref{tab:specfit}), and $\Delta t =$\xraydur\unskip\,(see Fig.~\ref{fig:lightcurves}) into Eq.~\ref{eq:xrayflare}. 
With Eq.~\ref{eq:ratio}, we find an energy of \eom\unskip\, for the corresponding OM flare. We use the $36\,$ks observing baseline of \textit{XMM-Newton} to calculate a flare rate in OM based on the detection of this single event. Fig.~\ref{fig:ffd} illustrates that the OM flare rate is roughly consistent, if slightly higher, than the extension of \textit{TESS}' FFD to lower energies. Assuming a 1:1 correspondence between optical and X-ray flares, about one flare per day with energies of the order of the observed one or above can therefore be expected in future X-ray observations of TIC~277, based on the \textit{TESS} FFD. However, usually, not all optical flares have X-ray counterparts~\citep{paudel2021simultaneous}, and those that do show order of magnitude variable ratios of optical vs. X-ray energy~\citep{guarcello2019simultaneous, kuznetsov2021stellar, joseph2024simultaneous}, implying that \textit{XMM-Newton} flare rates between $0.1-10\,$d$^-1$ are possible for TIC~277. 

The (marginal) OM flare precedes the soft X-ray flare, which is typical of flares that show the Neupert effect~\citep{neupert1968comparison}. This timing suggests that we observed the same event at two different wavelengths. 
The Neupert criterion states that the peak of the non-thermal radio emission should coincide with the moment of fastest rise in the thermal soft X-ray emission. As radio monitoring is rarely available for stellar flare observations, blue-optical and near ultraviolet have been used as proxies~\citep{hawley1995simultaneous, hawley2003multiwavelength}.
Many M dwarf flares follow the Neupert effect~\citep[e.g.][]{guedel1996neupert,stelzer2022great}, others deviate from it~\citep{hawley1995simultaneous,osten2005radio}, either not showing the time delay at all (non-Neupert flares), or a time delay inconsistent with the Neupert criterion~\citep[quasi-Neupert flares,][]{ tristan2023day}. In our data, the peak of the blue-optical flare peak precedes the X-ray peak by $12\,$min, overlapping with the fast rise phase of the soft X-ray flare. This suggests that this flare is a true Neupert flare, but the time resolution is too low to discriminate this instance from a potential quasi-Neupert case. Finally, we note that the X-ray flare light curve follows the typical fast-rise exponential decay shape, and is therefore unlikely to be eclipsed by the star despite its fast rotation. 

\subsection{Flare loop sizes and magnetic field strengths}
\label{sec:results:loops}

Based on the flare parameters obtained from EPIC, OM and \textit{TESS}, we estimated the flaring loop magnetic field strength $B$ and loop size $L$ using the scaling relations introduced in  Section~\ref{sec:methods:loops}.

For the EPIC flare, we use Eq.~\ref{eq:shibataB} to obtain the magnetic field strength, and Eq.~\ref{eq:shibataL} to derive the loop size from the flaring $EM$ and $T_{\rm hot}$, assuming flaring coronal densities $n_0=10^{11},10^{12}$, and $10^{13}\,\text{cm}^{-3}$~ (Table~\ref{tab:results:epicloop}), with the lower two values being typical of main-sequence M dwarfs with measured coronal species abundances~\citep{gudel2004xray, liefke2010multiwavelength}, and the latter only found in accreting T Tauri stars~\citep{stelzer2004xray}.

\begin{figure}
    \begin{centering}
        \includegraphics[width=\linewidth]{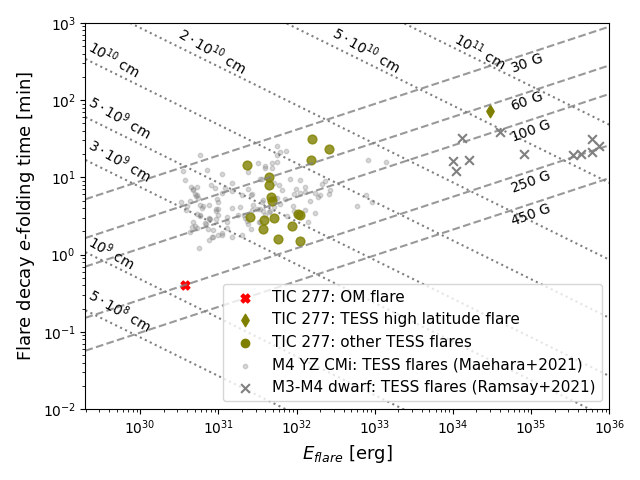}
        \caption{Flare decay \textit{e}-folding time $\tau$ vs. flare energy $E_{\rm flare}$ for the OM flare, and the \textit{TESS} flares. The dashed and dotted lines show the $\tau-E_{\rm flare}$ relations from Eqns.~\ref{eq:maeharaB} and \ref{eq:maeharaL}, for different loop magnetic field strengths and loop sizes, respectively. For comparison, we show the \textit{TESS} flares of the active M4 dwarf YZ CMi~\citep{maehara2021timeresolved} and a range of nearby M dwarfs from \textit{TESS} with large flares~\citep{ramsay2021transiting}.}
        
        \label{fig:maeharaBL}
    \end{centering}
\end{figure}

For the optical OM and \textit{TESS} flares, we follow~\citet{namekata2017validation, maehara2021timeresolved}, and use Eqns.~\ref{eq:maeharaB} and \ref{eq:maeharaL} to overlay the relation between \textit{e}-folding time $\tau$ and bolometric flare energy $E_{\rm flare}$ with a range of possible values for $B$ and $L$. We fit an exponential function to all \textit{TESS} flares and the OM flare, fixing the amplitude of the exponential decay to $5\%$ within the peak amplitude of the flare, and the start of the exponential decay to within $5\,$min or $20\,$s of the peak time of the flare in \textit{TESS} and OM, respectively. Figure~\ref{fig:maeharaBL} shows that $\tau$ and $E_{\rm flare}$ are correlated. The \textit{TESS} flares appear compatible with magnetic field strengths between $30\,$G and $250\,$G, similar to other active M dwarfs~\citep{karmakar2021astrosat, notsu2024apache}. Their loop sizes are between $3\cdot10^9\,$cm and $2\cdot10^{10}\,$cm, that is roughly $0.3-2\,R_*$, except for the energetic high-latitude flare, for which we estimated the energy and \textit{e}-folding time from the underlying, rotationally unmodulated flare~\citep{ilin2021altaipony}, and that yields a loop size of around $5\cdot10^{10}\,$cm, or $5\,R_*$. The OM flare extends the trend of the \textit{TESS} flares to lower energies with a magnetic field strength of $250\,$G and a loop size of $10^9\,$cm, or $0.1\,R_*$. This result is compatible with the results for the EPIC flare for loop number densities between $10^{11}\,\text{cm}^{-3}$ and $10^{12}\,\text{cm}^{-3}$~(see Table~\ref{tab:results:epicloop}). 

All scaling laws used here make assumptions about the state of the corona, e.g., the loop electron density~\citep{shibata2002hertzsprungrusselllike}, and the flare model, e.g., the magnetic reconnection model~\citep{namekata2017validation}. It is encouraging that the \textit{XMM-Newton} flare yields similar results for both EPIC and OM data, but the results for the \textit{TESS} flares should be interpreted cautiously, and rather differentially with other flare studies with the same method~\citep{maehara2021timeresolved} than as absolute values.

\begin{table}

    \centering
    \caption{Magnetic field strength $B$ and loop size $L$ of the EPIC flare, derived from Eqns.~\ref{eq:maeharaB} and \ref{eq:maeharaL}, respectively (see Section~\ref{sec:results:loops}).}
   \begin{tabular}{llll}

 $n_0 [\text{cm}^{-3}]$ & $10^{11}$ & $10^{12}$ & $10^{13}$ \\
\hline
$B$ [G] & 133 & 266 & 531 \\
$L$ [$R_*$] & 0.13 & 0.05 & 0.02 \\
\hline
\end{tabular}
    \label{tab:results:epicloop}
\end{table}

\section{Discussion}
\label{sec:discussion}

TIC~277's spectral type and rotation places it in the middle of the transition from star to brown dwarf. Our findings indicate that TIC~277's coronal temperature~(Section~\ref{sec:discussion:xraytemp}), flaring activity~(Section~\ref{sec:discussion:flares}), and flare loop sizes and magnetic fields~(Section~\ref{sec:discussion:loops}) are typical of saturated fully convective M dwarf stars. However, TIC~277 is rotating rapidly, even for the typically fast rotating late M dwarfs~\citep{medina2022galactic}. This combination could lead to an unusual manifestation of the stellar dynamo that produces high latitude flares~\citep{weber2016modeling, weber2023understanding}. TIC~277 is fainter in X-ray by about an order of magnitude than earlier-type M dwarfs on the canonical saturated branch~\citep{magaudda2022firsta}, which, in combination with the detection of a high latitude flare, may be an indication of polar updraft migration that is suspected to occur in rapid rotators~(Section~\ref{sec:discussion:xraylum}).

\subsection{Coronal temperature}
\label{sec:discussion:xraytemp}

Low mass stars' coronae are not uniform. Magnetic structures create a range of temperatures, starting at about 1 MK, and rising up to 30 MK in flares, in the heterogeneous Solar corona~\citep{vaiana1978recent}. Stellar point source X-ray observations are a superposition of various temperatures and densities. The resolution of different temperature components depends on the apparent brightness and integration time of the spectrum. Nearby stars like Proxima Cen can be modeled with a range of temperatures~\citep[e.g.,][]{gudel2004flares, drake2020pointing}, while one can often only capture a single or a few dominant temperatures in fainter objects.  

We find that the \textit{XMM-Newton} observations of the quiescent corona of TIC~277 are best described by two temperature components, similar to many other active M dwarfs. Assuming that TIC~277 produces flares below the detection threshold of \textit{XMM-Newton} that
contribute to the temperature makeup of the quiescent corona, the presence of a second, hot component is consistent with the standard picture for stellar flares~\citep{wargelin2008xray, robrade2010quiescent, behr2023muscles, magaudda2022firsta}. Both the cooler and the hotter component represent thermal radiation from hot plasma that evaporates from the chromosphere into the corona. The dominance of the hot component during the flare combined with the observed Neupert effect (see Section~\ref{sec:results:flares}) is evidence that this component is driven by particles precipitating downward during flares that then heat the chromosphere to high temperatures, and cause evaporation into the coronal flaring loops. In addition, quiescent heating through other sources under debate, such as Alfv\'en waves or nanoflares~\citep{benz2016flare}, are present at a constant level represented by the cool component.

The range of $6.5-11\,$MK for the emission measure weighted temperature of TIC~277 places it squarely in the saturated regime with other mid M dwarfs down to spectral type M6~\citep{wright2018stellar, magaudda2020relation, stelzer2022first,robrade2005xray,raassen2003xray,paudel2021simultaneous, foster2020corona}. 
Flares on the slowly rotating Proxima Cen (M5.5, $P_{\rm rot} = 83\,$d,~\citealt{anglada-escude2016terrestrial}), appear with a range of temperatures at 2-3 MK and 10-20 MK~\citep{gudel2004flares, fuhrmeister2011multiwavelength, fuhrmeister2022high, howard2022mouse}. Its X-ray luminosity~\citep{haisch1990rotational} and flaring activity~\citep{howard2018first, medina2020flare} place it in the transition region between saturated and unsaturated regimes. Only at rotation periods beyond 100 days, stellar activity declines to yield coronal temperatures $<2\,$MK in M3-M6 dwarfs~\citep{wright2018stellar, foster2020corona}. 

There are only few late M dwarfs beyond spectral type M6 with a spectrally resolved corona. Their intrinsic faintness and gradual decline in coronal emission toward the brown dwarf regime make them hard to detect~\citep{berger2010simultaneous, cook2014trends, stelzer2022first}. One example is NLTT 33370 AB. This M7 binary with a rotation period of about $3.8\,$h (close to TIC~277's $4.56\,$h, and of the same spectral type) is composed of a 3.1 MK and 14 MK component~\citep{williams2015simultaneous}. Another is TRAPPIST-1~(M8, \mbox{$P_{\rm rot} = 3.3\,$d},~\citealt{luger2017sevenplanet}). Its corona appears cooler, with a 1.74 MK and a 9.6 MK component~\citep{wheatley2017strong}, and an emission measure weighted temperature of 5.36 MK~\citep{brown2023coronal}. 

Bearing in mind the relatively low two-temperature resolution of TIC~277's X-ray spectrum and the dearth of X-ray spectra for stars at the bottom of the main sequence, it shows a coronal temperature make-up common for mid-to-late M dwarfs in the saturated activity regime, which it will possibly keep for gigayears until it spins down to very low rotation rates~\citep{medina2022galactic, engle2023living}. 

\subsection{Flaring activity}
\label{sec:discussion:flares}

TIC~277's flare rate is consistent with other saturated fully convective dwarfs~\citep{medina2020flare, murray2022study}, and so is the energy distribution with a power law slope of \ffdalpha\unskip. It is in fact comparable to TRAPPIST-1's flaring behavior~\citep{paudel2018k2}, which has been investigated for its effects on the seven terrestrial planets in its orbit. 


TIC~277's rapid rotation, and UV flux hint at a younger age than TRAPPIST-1, which is likely a field star~\citep{burgasser2017age, birky2021improved,gonzales2019reanalysis}. 
TIC~277 has not been attributed to any moving group so far~\citep{schneider2018hazmat}, and the number of late M dwarfs with independently measured age and rotation is too small to infer age based on rotation alone~\citep{engle2023living}. Interpreting the upcoming observations of TRAPPIST-1 e, one of the rocky planets in the habitable zone of the system, with the James Webb Space Telescope will most likely involve its activity history. Studies of its younger counterparts, late M dwarfs like TIC~277, are required to empirically constrain cumulative effects of atmospheric forcing of planets. If its energetic flares commonly occur at high latitudes~\citep{ilin2021giant}, future models of energetic particle exposure of habitable zone planets may have to include age dependent latitudes of particle eruption to reproduce observations. 

\subsection{Flare loop sizes and magnetic field strength}
\label{sec:discussion:loops}
The \textit{XMM-Newton} flare emitted an energy of \eepic, comparable to that of energetic solar flares, and the loop size derived from both EPIC and OM is within their range of $10^{8} -10^{10}\,$cm~\citep{aschwanden2015global, namekata2017statistical}. The \textit{TESS} flares show loop sizes between 0.3 and 5 $R_*$, consistent with M dwarf flares with similar energies~\citep{schmitt2002xray, stelzer2006simultaneous, karmakar2021astrosat, stelzer2022great, ramsay2021transiting, notsu2024apache, guarcello2019simultaneous, doyle2022doubling}.

The \textit{XMM-Newton} flare observations suggest a magnetic field strength around $150-250\,$G from both the EPIC and OM data, stronger than the fields of solar coronal loops which are typically below $100\,$G~\citep{namekata2017statistical}. However, solar flares have shown higher field strengths up to $350\,$G in high spatial resolution spectropolarimetric and radio observations, suggesting that lower resolution underestimates the field strength~\citep{kuridze2019mapping, yu2020magnetic}. Since the scaling relations used in this work are based on the older solar flare measurements, the estimates we derived may be too low by up to a factor of five~\citep{kuridze2019mapping}.

In a differential comparison, the magnetic field strengths of $30-250\,$G of the \textit{TESS} flares follow the trend of other activity saturated early to mid M dwarfs~(Fig.~\ref{fig:maeharaBL},~\citealt{ramsay2021transiting, maehara2021timeresolved}). For late M dwarfs like TIC~277, only few estimates of flare loop sizes and field strengths exist in the literature~\citep{schmitt2002xray, stelzer2006simultaneous}, which are consistent with our results, but there are no such estimates for flare energies below $10^{31}$\,erg. 

Overall, TIC~277's flare loop sizes and field strengths appear consistent with those of active M dwarfs of similar energies. However, we cannot say if other late M dwarfs behave similarly in the low energy range of the \textit{XMM-Newton} flare. Late M dwarf flares at energies below $10^{31}\,$erg have been reported~\citep[e.g.,][]{paudel2018k2, howard2022mouse, petrucci2024exploring}, but we leave the worthwhile exercise of estimating their loop lengths and field strengths to future work. 


\subsection{X-ray luminosity}
\label{sec:discussion:xraylum}


TIC~277's coronal luminosity relative to bolometric luminosity of $L_X/L_{\rm bol}=$\LXLbol in quiescence is about an order of magnitude lower than the canonical $L_X/L_{\rm bol}\sim 10^{-3}$ average in the saturated regime of partly and fully convective M dwarfs~\citep{wright2011stellaractivityrotation,wright2016solartype,wright2018stellar}.
However, recent studies limited to low-mass M dwarfs ($M_* \sim 0.15-0.3\,M_\odot$) have shown that the saturation level is near  $L_X/L_{\rm bol} \approx 10^{-3.5}$~\citep{magaudda2022first} 
On the contrary, in a sample of late-M dwarfs investigated with eROSITA $L_X/L_{\rm bol}$ appears to increase towards later spectral types \citep{stelzer2022first}. However, these detections are likely to be attributed to flares and, thus, not representative for the average X-ray emission level of late-M dwarfs. While TIC\,277 appears at the lower end of the $L_X/L_{\rm bol}$ distribution with respect to that sample, and similar previous work \citep{stelzer2012ultracool, cook2014trends, deluca2020extras, williams2014trends, berger2008simultaneous}, the latest eROSITA study on a well-defined sample of nearby late-M to early-L dwarfs has shown 
that rapidly rotating late M dwarfs display a broad range of X-ray luminosities centered around $L_X/L_{\rm bol}\approx 10^{-4}$ placing TIC\,277 in the center of their distribution~\citep{magaudda2024transitions}.


At late-M spectral types, a decline in coronal activity is expected as the magnetosphere transitions from a stellar to a planetary one~\citep{pineda2017panchromatic}. 
A decline in X-ray activity is expected as a result of decreasing efficiency of coronal heating~(e.g.,~\citealt{mohanty2002activity, williams2014trends}).  However, the flaring activity and coronal temperature of TIC~277 
and similar late-M dwarfs show that the corona is still capable of producing flares.
For the specific case of the rapidly rotating TIC~277,
 polar updraft migration~\citep{stepien2001rosat} could at the same time explain the decline in X-ray luminosity and the occurrence of a high latitude flare observed by \textit{TESS}. Polar updraft migration implies that at high rotation rates, flux emergence becomes more efficient near the poles than at the equator~\citep{yadav2015formation, weber2016modeling} producing flares there, but draining the equatorial regions of magnetic flux required to produce a corona at low latitudes. As a result, the X-ray luminosity diminishes. Since TIC~277 is seen nearly equator-on~(Table~\ref{tab:starparams}), polar updraft migration could explain its fainter corona. Considering the high-latitude flare, viewing geometry, star-like flaring behavior and star-like coronal temperatures together, we suggest this scenario as an alternative to the brown dwarf transition explanation, that would invoke a diminishing corona due to the overall cooler, more neutral atmosphere. 

\section{Summary and Conclusions}
\label{sec:summary}
We investigated whether TIC~277, a rapidly rotating M7 dwarf that exhibited a flare localized at $81^{\circ}$ latitude, shows coronal or flaring properties that could explain the occurrence of this flare so close to the rotational pole. We obtained 36 ks of EPIC and OM observations with \textit{XMM-Newton}, and studied the five Sectors of red-optical 27-day light curves provided by \textit{TESS}. We found a mean quiescent coronal temperature of \Tqmean, and a flaring rate $f(> \log_{10} E_{\rm flare, bol} = 31.5\,\rm{erg})\approx$\ffdr with an energy distribution with a slope of \ffdalpha, all typical of M dwarfs in the saturated regime. We also detected a simultaneous X-ray and optical flare with an energy of \eepic\,in the $0.2-2\,$keV band, and bolometric energy of \eom\,from the OM observations. In combination with the 18 detected \textit{TESS} flares, we estimate that X-ray flares on TIC~277 can be observed with \textit{XMM-Newton} at least about once a day. We used scaling relations to estimate the loop sizes and magnetic fields of the flares, and found mutually consistent fields of $30-250\,$G and loop lengths ranging from $0.1R_*$ for the low energy \textit{XMM-Newton} flare to $5R_*$ for the giant, high latitude \textit{TESS} flare. Similar values have been reported for activity saturated M dwarfs in the literature, except for the \textit{XMM-Newton} flare, which extends the duration-energy relation to more impulsive and less energetic flares. 
An indication of an unusual corona stems from its X-ray luminosity relative to bolometric luminosity $L_X/L_{\rm bol}=$\LXLbol\hspace{-0.1cm}. It is an order of magnitude lower than the canonical value for saturated activity M dwarfs, but representative of late M dwarfs of spectral type M7 and later. TIC~277 is an extremely rapid rotator with a rotation period of only $4.56\,$h. The detected high-latitude flare in \textit{TESS}~\citep{ilin2021giant} may hence be a product of high-latitude flux emergence driven by its fast rotation~\citep{weber2016modeling,weber2017suppression}. The low X-ray luminosity and the high-latitude flare taken together with the star's fast rotation, saturated flaring activity, and equator-on viewing angle could be explained by polar updraft migration, which suppresses coronal emission in equatorial regions, and confines the corona and flaring regions to near-polar latitudes~\citep{stepien2001rosat}. 

Overall, less than ten flare latitudes on stars other than the Sun are known to date~\citep{wolter2008doppler, ilin2021giant, johnson2021simultaneous}. Localization of more flares on the surfaces of a broader range of stars will allow us to resolve if polar updraft migration indeed occurs in these stars.

\section*{Acknowledgements}
 E.I. and D.D. acknowledge funding from the Deutsche Luft- und Raumfahrtgesellschaft (FKZ 50 OR 2209, project \textit{X-ray Loops}). K.P. acknowledges funding from the German \textit{Leibniz-Gemeinschaft} under project number P67/2018. This work is based on observations obtained with \textit{XMM-Newton}, an ESA science mission with instruments and contributions directly funded by ESA Member States and NASA. This paper includes data collected by the \textit{TESS} mission. Funding for the \textit{TESS} mission is provided by the NASA's Science Mission Directorate. This work made use of the open source python software packages \texttt{lightkurve}~\citep{lightkurvecollaboration2018lightkurve}, \texttt{astropy}~\citep{robitaille2013astropy}, \texttt{numpy}~\citep{harris2020array}, \texttt{pandas}~\citep{reback2022pandasdev}, \texttt{matplotlib}~\citep{hunter2007matplotlib}, \texttt{emcee}~\citep{foreman-mackey2013emcee}, \texttt{scipy}~\citep{mckinney2010data}, and \texttt{altaipony}~\citep{ilin2021altaipony}. This work has made use of data from the European Space Agency (ESA) mission
{\it Gaia} (\url{https://www.cosmos.esa.int/gaia}), processed by the {\it Gaia}
Data Processing and Analysis Consortium (DPAC,
\url{https://www.cosmos.esa.int/web/gaia/dpac/consortium}). Funding for the DPAC
has been provided by national institutions, in particular the institutions
participating in the {\it Gaia} Multilateral Agreement.

\section*{Data Availability}
All data used in this study are publicly available in their respective archives, i.e., the \href{https://mast.stsci.edu/portal/Mashup/Clients/Mast/Portal.html}{Mikulski Archive for Space Telescopes (MAST)} for \textit{TESS}, and the \href{https://www.cosmos.esa.int/web/xmm-newton/xsa}{\textit{XMM-Newton} Science Archive}.
The scripts used to produce the figures in the git repository on GitHub are available at:~\href{https://github.com/ekaterinailin/tic277-paper}{github.com/ekaterinailin/tic277-paper}. The data analysis scripts can be found under \href{https://github.com/ekaterinailin/tic277}{github.com/ekaterinailin/tic277}.

\bibliographystyle{aa} 
\bibliography{references}

\end{document}